# Band alignment at epitaxial BaSnO$_3$/SrTiO$_3$(001) and BaSnO$_3$/LaAlO$_3$(001) heterojunctions

Scott A. Chambers[1,*], Tiffany C. Kaspar[1], Abhinav Prakash[2], Greg Haugstad[3], Bharat Jalan[2,*]

[1]*Physical and Computational Sciences Directorate, Pacific Northwest National Laboratory, Richland, WA 99352*
[2]*Department of Chemical Engineering and Materials Science, University of Minnesota, Minneapolis, MN 55455*
[3]*Characterization Facility, University of Minnesota, Minneapolis, MN 55455*

We have spectroscopically determined the band offsets at epitaxial interfaces of BaSnO$_3$ with SrTiO$_3$(001) and LaAlO$_3$(001). The conduction band minimum is lower in electron energy in the BaSnO$_3$ than in the SrTiO$_3$ and LaAlO$_3$ by 0.6 ± 0.1 eV and 3.7 ± 0.1 eV, respectively. Thus, electrons generated in the SrTiO$_3$ and LaAlO$_3$ and transferred to the BaSnO$_3$ by modulation and polarization doping, respectively, are expected to drift under the influence of an electric field without undergoing impurity scattering and the associated loss of mobility. This result bodes well for the realization of oxide-based, high-mobility, two-dimensional electron systems that can operate at ambient temperature.

*Authors to whom correspondence should be addressed. Electronic mail: sa.chambers@pnnl.gov and bjalan@umn.edu

BaSnO$_3$ (BSO) is an attractive wide-bandgap semiconductor in the field of oxide electronics. When doped *n*-type, BSO exhibits considerably higher room-temperature electron mobility than SrTiO$_3$ (STO), which has traditionally been taken to be the leading perovskite oxide semiconductor. Bulk single crystals of BSO have been shown to exhibit room-temperature mobility values as high as 320 cm$^2$/V-s,[1] whereas epitaxial films typically yield lower values thus far, ranging between 10 and 150 cm$^2$/V-s.[2-6] Additionally, BSO has high optical transparency, rendering it of considerable interest for transparent electronic technologies.[7-10] While there is potential to increase room-temperature electron mobility in doped-BSO films through defect minimization, heterostructure engineering also provides an exciting route to this end. Heterostructure engineering provides pathways to isolate the carriers from scattering centers such as the dopants from whence they come. This can be done by either modulation doping (introducing dopants in a different material across an interface from BSO) or polarization doping (inducing charge transfer into non-polar BSO via interface formation with a polar material). In either scheme, the conduction band minimum in the BSO film should be of lower electron energy than that of the material to which the BSO is joined in order to confine carriers in the BSO. Under this condition, free carriers will readily spill over into and remain in the BSO layer, provided the conduction band offset is sufficiently large. STO is a good choice as an electron source for the modulation doping scheme, and LaAlO$_3$ (LAO) is suitable to test the polarization doping approach.

To these ends, we have deposited epitaxial undoped BSO films on undoped STO(001) and LAO(001) substrates using hybrid molecular beam epitaxy, and have measured the band offsets and BSO band gap using *ex situ* x-ray photoelectron spectroscopy (XPS) and spectroscopic ellipsometry (SE), respectively. Details of our hybrid MBE method for BaSnO$_3$ heteroepitaxy are discussed elsewhere.[11] Film compositions were determined using Rutherford backscattering spectrometry (RBS) and SIMNRA spectral simulation software.[12]

High-resolution core-level (CL) and valence band (VB) x-ray photoelectron spectra were acquired *ex situ* using monochromatic Al K$\alpha$ x-rays ($h\nu$ = 1486.6 eV) and a VG/Scienta R3000 electron energy analyzer. The specimens were exposed to a 15 minute UV/ozone treatment immediately prior to insertion into the XPS system load lock to minimize the concentration of adventitious carbon on the surface. All spectra were measured in the normal-emission geometry and with a total energy resolution of 0.45 eV. A low-energy electron flood gun was required to neutralize the positive photoemission charge that builds up on the surface of these insulating specimens. SE was carried out using a J.A. Wollam V-VASE ellipsometer. The optical properties of the BSO films were modeled as a series of Gaussian (below the bandgap) and Lorentzian (above the bandgap) oscillators; the primary absorption edge was modeled with a Tauc-Lorentz function..

Figures 1a&b show out-of-plane wide-angle x-ray diffraction (XRD) scans for 28±1 unit cell (u.c.) BSO films on STO and LAO (001), respectively. A single Bragg peak is observed, indicating phase purity, for BSO films on both substrates for which the in-plane lattice mismatches ($\Delta a/a_{BSO}$) are -5.1 % (STO) and -7.9 % (LAO). Finite thickness fringes are also evident, indicating excellent crystallinity, as well as smooth surfaces and buried interfaces. Contact mode AFM images of the film surfaces are shown as insets. These images reveal atomically smooth surface morphology for both films, thereby corroborating the XRD results. The *c* lattice parameters were determined to be 4.137(5) Å and 4.122(5) Å for films on

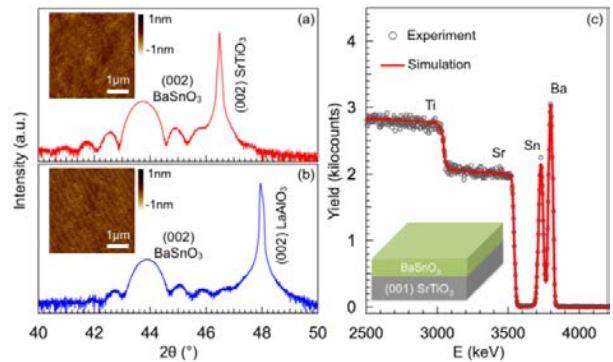

Figure 1. Out-of-plane XRD scans for 28 u.c BSO on (a) STO(001) and (b) LAO(001), along with AFM images of the film surfaces as insets. (c) RBS (He$^{++}$ at 4.3 MeV) and SIMNRA simulation for a 134 u.c BSO/STO(001) specimen prepared under conditions identical to those used for growing the 5 and 28 u.c. films utilized to determine the band offsets.

STO and LAO, respectively. These values are larger than the bulk value of 4.116 Å, which we attribute to incomplete strain relaxation in the 28 u.c. films. A much thicker (82 u.c.) film on LAO(001) grown under same conditions yielded lattice parameters identical to those of bulk BSO, suggesting that the expanded *c* lattice parameters in our 28 u.c. films are not due to non-stoichiometry in the cations or oxygen vacancy defects, but rather to incomplete strain relaxation.[11] We show in Figure 1c RBS data for a representative 134 u.c. BSO film on STO(001), along with a simulation yielding a Ba/Sn ratio of 0.996±0.010. The agreement is very good, providing direct evidence for nearly stoichiometric BSO.

We use a straightforward method for measuring valence band offsets (VBO) in which the valence band maximum (VBM) for each material in a heterostructure is referenced to select CL photoelectron peaks uniquely associated with that material.[13-18] The binding energy differences between appropriate CL peaks are then used to determine the VBO. Figure 2 shows CL and VB spectra for the 28 u.c. films on STO and LAO. For these spectra, the binding energy scale is referenced to the VBM of each film. The Ba 4d spin-orbit peaks are asymmetric, and require a second pair of peaks shifted ~1 eV to higher binding energy in order to obtain a good fit. We assign the features at higher binding energy to Ba ions at the surface that are either oxidized to form $BaO_2$[19] or are hydroxylated upon air exposure, while the more intense peaks are associated with subsurface lattice Ba. From these spectra, we obtain the energy differences between select core peaks (Sn $3d_{5/2}$ and lattice Ba $4d_{5/2}$ indicated by arrows in Fig. 2) and the associated VBM. These numbers are required to obtain the VBOs from CL spectra for heterojunctions made from thinner (5 u.c.) BSO films.

Figure 3 shows the core peaks used to obtain the VB offsets for heterojunctions of 5 u.c. BSO on STO and LAO. This BSO thickness was chosen to be large enough that the electronic structure of the BSO is likely to be fully developed, yet thin enough to yield sufficient photoemission signal from the substrate core peaks that their binding energies can be accurately measured. The Ba 4d peaks show a larger intensity fraction at the higher binding energies than do the spectra for the 28 u.c. films, presumably because the number of near-surface layers prone to chemical reaction with the ambient is a larger fraction of the total number of layers in the former than in the later. The La 4d spectrum associated with the BSO/LAO system also shows considerable structure, but this is due to shake-up rather than surface over-oxidation. The high degree of symmetry in the Sn $3d_{5/2}$ peaks is consistent with the undoped character of the films.[4] For both heterostructures, we use different pairs of CLs in order to generate independent values of the VBO for each heterojunction.[20,21] In the case of BSO/STO, we use: (i) Sn $3d_{5/2}$ & Ti $2p_{3/2}$, and, (ii) Ba $4d_{5/2}$ & Sr $3d_{5/2}$, whereas for BSO/LAO, we use: (iii) Ba $4d_{5/2}$ & La $4d_{5/2}$, and, (iv) Ba $4d_{5/2}$ & Al 2p. The band offsets can then be determined using the following formulae:

$$\Delta E_V = (E_{Sn3d5/2} - E_V)_{BSO} - [(E_{Sn3d5/2} - E_{Ti2p3/2})_{HJ} + (E_{Ti2p3/2} - E_V)_{STO}]$$

$$\Delta E_V = [(E_{Ba4d5/2} - E_V)_{BSO} + (E_{Sr3d5/2} - E_{Ba4d5/2})_{HJ}] - (E_{Sr3d5/2} - E_V)_{STO}$$

$$\Delta E_V = [(E_{Ba4d5/2} - E_V)_{BSO} + (E_{La4d5/2} - E_{Ba4d5/2})_{HJ}] - (E_{La4d5/2} - E_V)_{LAO}$$

$$\Delta E_V = (E_{Ba4d5/2} - E_V)_{BSO} - [(E_{Ba4d5/2} - E_{Al2p})_{HJ} + (E_{Al2p} - E_V)_{LAO}]$$

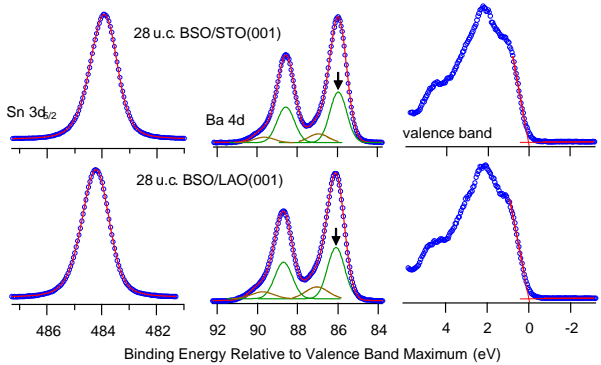

Figure 2. Sn $3d_{5/2}$ and Ba 4d core-level x-ray photoelectron spectra, along with x-ray excited valence band spectra for 28 u.c BSO/STO(001) and BSO/LAO(001) heterostructures.

Inspection of these formulae shows that in all cases, the sign convention is such that the VB offset (VBO) will be negative if the VBM for the BSO is at a *higher* photoelectron binding energy (lower electron energy) than that for the substrate.

We show in Table 1 the core-to-VBM values for the 28 u.c. films along with VBO values for the thin-film heterojunctions. There is a non-negligible difference between the Sn $3d_{5/2}$-to-VBM energy splittings for the thicker films on LAO and STO. This difference is most likely an artifact of the charge neutralization process required to measure high-resolution XPS on insulating samples such as these. Unexpected differences are also observed in the energy splittings between the Ba 4d and Sn 3d peaks between thick and thin films, and in the energy splittings between Sr 3d and Ti 2p peaks for thin BSO/STO and bulk STO. In order to minimize the effect of these differences on the VBO determination, the two peaks in each CL pair were chosen to be of comparable binding energy (e.g. either in the 450 – 490 eV range, the 80 – 130 eV range, or the 70 – 110 eV range). With these choices, the VB offsets from the different CL pairs yield consistent results, as seen in Table 1. For heterojunctions on both STO and LAO, the BSO VBM is at lower electron energy than that of the substrate by a few tenths of an eV. As an independent test of the CL method, and to be sure that the use of higher binding energy cores in the case of BSO/STO isn't skewing the VBO results, we verified the VBO directly using VB spectra. We did so by simulating the VB for the 5 u.c. BSO/STO heterojunction by taking a linear combination of spectra for the STO substrate and the thick BSO film after weighting the intensities to account for film thickness, and shifting the BSO spectrum 0.25 eV (the average VBO from the two CL pairs) to higher binding energy. This simulated spectrum matches well the measured spectrum with regard to overall shape and width, as seen in Figure 5, thereby corroborating the CL result.

Table 1 – Binding energy differences and valence band offsets (in eV) for BSO/STO(001) and BSO/LAO(001) heterojunctions

|  | $E_{Ba4d} - E_V$ | $E_{Sn3d} - E_V$ | $\Delta E_V$ | $\Delta E_V$ |
|---|---|---|---|---|
| 28 uc/STO | 85.95(4) | 483.88(2) | ----- | ----- |
| 28 uc/LAO | 86.10(4) | 484.16(2) | ----- | ----- |
| 5 uc/STO | ----- | ----- | -0.18(4) (i) | -0.32(6) (ii) |
| 5 uc/LAO | ----- | ----- | -0.37(8) (iii) | -0.50(8) (iv) |

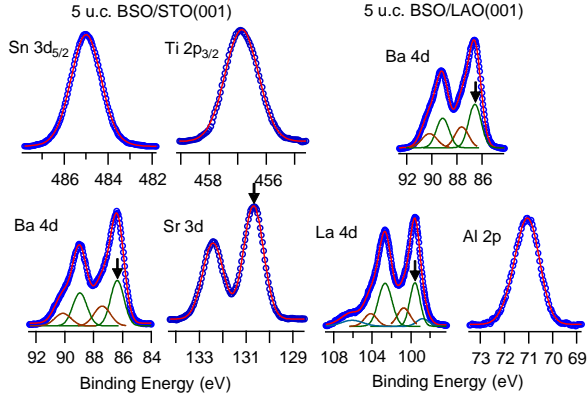

Figure 3. Core-level x-ray photoelectron spectra for 5 u.c BSO/STO(001) and 5 u.c. BSO/LAO(001) heterostrutures.

In order to obtain the CB offset (CBO), we need an accurate value of the indirect band gap for BSO. To this end, we used SE to measure the indirect and direct gaps for the 28 u.c. BSO films on LAO and STO. Figure 4a shows the refractive indices ($n$) and extinction coefficients ($k$) for the two films. These results are in reasonable agreement with the dielectric function measured by Luo et al.[8] on a single-crystal BSO specimen by SE. (Note that $(n + ik)^2 = \varepsilon_1 + i\varepsilon_2$, where $\varepsilon_1$ and $\varepsilon_2$ are the real and imaginary components of the dielectric function, respectively.) The absorption coefficients ($\alpha$) for the two films are shown in Figure 4b. The indirect gaps were determined by constructing Tauc plots of $(\alpha h\nu)^{1/2}$ vs $h\nu$ with $\alpha = 4\pi k/\lambda$,[22] and the resulting indirect gaps are 2.89(5) eV and 2.95(5) eV for growth on STO and LAO, respectively, as seen in Figure 4c&d. The associated direct gaps were found to be 3.55(5) eV for both films from Tauc plots of $(\alpha h\nu)^2$ vs $h\nu$.[23] Our indirect gap values are in good agreement with those reported by Lebens-Higgins et al.[4] based on hard x-ray XPS at $h\nu = 4$ keV. These authors measured the difference between the VBM and CBM to be ~3.07 eV for La-doped BSO epitaxial films with very low doping levels grown by MBE. Likewise, Kim et al.[24] reported indirect and direct bandgap values of 2.95 and 3.10 eV, respectively, for an undoped single-crystal BSO specimen. The indirect bandgap values from Figures 4c&d agree well with the value reported by Kim et al.[24] The optical and electronic properties in Fig. 4 are also in quantitative agreement with those obtained by recent density functional theory (DFT) calculations using the modified Becke-Johnson type potential functional of Tran and Blaha, which predicted an indirect bandgap of 2.82 eV, an index of refraction in the range of ~1.85 – 2, and a similar dependence of the absorption coefficient on photon energy for unstrained, undoped BSO.[9,10] In contrast, the experimental bandgap values reported for BSO powder samples range from 3.1 eV[25] to 3.4 eV,[26] likely due to a strong Burstein-Moss shift arising from unintentional defect- or impurity-induced carrier doping. The lack of a Burstein-Moss shift in the bandgap values reported here confirms the insulating nature of the thick epitaxial films.

Having determined the indirect bandgap for BSO, the CBO is readily determined from the VBO and the difference in indirect gap values for the film and substrate materials as $\Delta E_C = \Delta E_g - \Delta E_V$. Again, the sign convention is such that $\Delta E_C$ is negative if the CB minimum (CBM) in the BSO is at a lower electron energy than that of the substrate. The resulting CBOs, along with the VBOs and indirect gaps, are summarized in Figure 5. Here we average over the two CL pairs summarized in Table I to obtain the VBO values for each kind of heterojunction. Our VBO for BSO/STO (-0.25 eV) is in excellent agreement with that calculated from DFT using the HSE06 hybrid functional by Krishnaswamy et al.[27] (-0.27 eV). However, these authors also calculate an indirect gap of 2.40 eV, which differs significantly from our measured value of ~2.9 eV. This discrepancy propagates into a similar discrepancy in the CBO between theory (-1.14 eV, using our sign convention), and experiment (-0.6 eV). Because of the relatively low CB density of states these authors calculate for BSO, the CBO we measure (-0.6 eV) limits the carrier density that can be confined within BSO films using a modulation doped structure with STO to ~ 4-6 ×10$^{13}$ cm$^{-2}$. However, with its larger CBO (-3.7 eV), the BSO/LAO heterostructure should

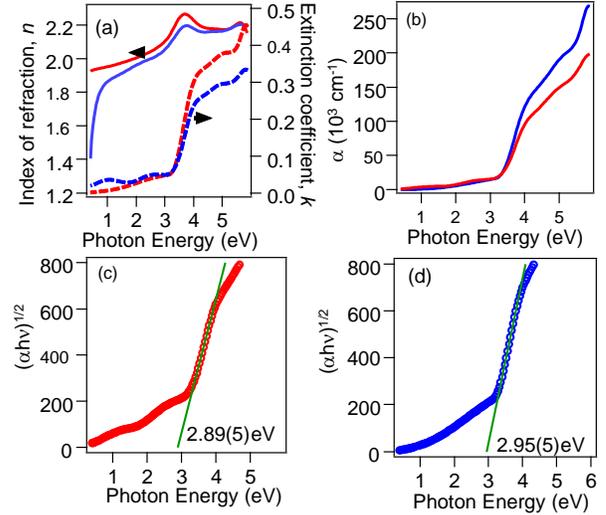

Figure 4. (a) Indices of refraction and extinction coefficients and (b) absorption coefficients measured by spectroscopic ellipsometry for 28 u.c BSO/STO(001) (red) and 28 u.c. BSO/LAO(001) (blue). Tauc plots with exponent ½ yielding indirect gaps values for (c) 28 u.c BSO/STO(001) and for (d) 28 u.c. BSO/LAO(001).

provide better carrier confinement up to ~ 3-4 ×10$^{14}$ cm$^{-2}$.[27]

The experimental CBOs we report reveal that for both modulation doping with STO and polarization doping with LAO, the electrons should readily transfer into the BSO, thereby separating from their respective sources. The resulting heterostructures should thus exhibit higher mobilities than would single films of La-doped BSO. Because $\Delta E_C$ is negative, the BSO/STO and/or BSO/LAO interfaces may facilitate low-density, high-mobility electron gases, thereby paving the way for the investigation of integer/fractional quantum Hall effects. In addition, having a high mobility channel at room temperature in an engineered complex oxide heterostructure may enable oxide-based, high-mobility, two-dimensional electron systems that can operate at ambient temperature. Such a material system would be highly useful for all-perovskite transistors in power electronics applications.

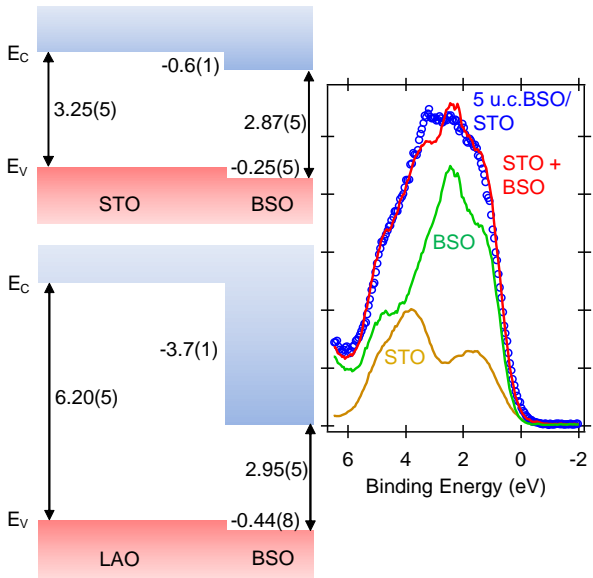

Figure 5. Energy diagrams showing band alignment at the BSO/STO and BSO/LAO heterojunctions. VB spectrum for the 5 u.c. BSO/STO system (blue circles), along with a simulation (red) done by taking a linear combination of bulk STO (brown) and 28 u.c. thick BSO film (green) spectra after shifting the BSO spectrum 0.25 eV to higher binding energy, and weighting to account for the BSO film thickness.

The thin film growth and characterization work at the University of Minnesota was supported primarily by the National Science Foundation through DMR-1410888 and in part by the MRSEC under award # DMR-1420013. We also acknowledge use of facilities at the UMN Minnesota Nano Center. Parts of this work were carried out in the Characterization Facility, University of Minnesota, which receives partial support from NSF through the MRSEC program. The XPS and SE work at PNNL was supported by the U.S. Department of Energy, Office of Science, Division of Materials Sciences and Engineering under Award #10122. The PNNL work was performed in the Environmental Molecular Sciences Laboratory, a national scientific user facility sponsored by the Department of Energy's Office of Biological and Environmental Research and located at PNNL.